\documentclass[twocolumn,showpacs,amsmath,amssymb,amsfonts,nofootinbib,prl]{revtex4}
\usepackage{graphicx}
\usepackage{dcolumn}
\usepackage{bm}
\usepackage{ulem}

\def\beq{\begin{equation}}
\def\eeq{\end{equation}}

\begin{document}
\input{epsf}

\title{Generalisation of Gunion-Bertsch Formula for Soft Gluon Emission}

\author{Raktim Abir$^1$, Carsten Greiner$^2$, Mauricio Martinez$^{2,3}$, 
and Munshi G. Mustafa$^{1}$}
\affiliation{$^1$Theory Division, Saha Institute of Nuclear Physics  
1/AF Bidhannagar, Kolkata 700064, India.}

\affiliation{$^2$Institute f\"ur Theoretische Physik, Johann Wolfgang Goethe
University, Max-von-Laue-Strasee 1, D-60438 Frankfurt, Germany} 

\affiliation{$^3$Frankfurt Institute for Advanced Studies, Ruth-Moufang-Strasse 1,
D-60438, Frankfurt, Germany}

\begin{abstract}
We generalise the most extensively used Gunion-Bertsch formula 
for the soft gluon emission in a perturbative QCD. We show that
the corrections arising due to this generalisation  could be  
very important for the phenomenology of the hot and dense matter produced
in the heavy-ion collisions. 
\end{abstract}


\date{\today}
\maketitle


The prime intention for ultra relativistic heavy-ion collisions is to study
the behaviour of nuclear or hadronic matter at extreme conditions like very
high temperatures and energy densities. A particular goal lies in the
identification of a new state of matter formed in such collisions, the
quark-gluon plasma (QGP), where the quarks and gluons are deliberated from
the nucleons and move freely over an extended space-time region. Various
measurements taken in CERN-SPS~\cite{heinz} and
BNL-RHIC~\cite{white,dilep,phot,ellip,jet,phenix} do lead to 'wealth of 
information' for the formation of QGP through the hadronic final states. 
In the upcoming experiments at the CERN LHC, one is hoping to produce 
QGP during the first several fm/$c$ of the collisions and to substantiate 
those evidences already found in the past as well in the recent experiments.

Some quantitative key features of the plasma produced in such collisions
include equilibration and its time, initial temperature, energy-loss
and jet quenching of high energetic partons, and elliptic flow of hadrons
and its scaling with the number of valence quarks. 
The Gunion-Bertsch (GB) formula~\cite{gunion82} for soft gluon emission has 
widely been used for various aspect of the heavy-ion phenomenology. 
To set the perspective we note that there are many papers 
in the literature based on the GB formula. We recall: 
the sequence of events in hot glue scenario~\cite{Baier:2000sb,asr94,xiong94,
El:2007vg}, 
thermal equilibration~\cite{Xu:2004mz,Xu:2007aa},  
gluon chemical equilibration~\cite{biro93,mgm97,Xu:2004mz},
parton matter viscosity~\cite{Xu:2007jv,Chen:2009sm},
radiative energy-loss of high energy partons propagating through a 
thermalised QGP~\cite{Fochler:2008ts,Fochler:2010wn,mgm08} etc., where 
the GB formula has extensively been used. The original GB formula was 
derived in Ref.\cite{gunion82} for gluon emission 
from quark-quark scattering and later it was explicitly used in 
Ref.\cite{biro93,wong96} to derive the soft gluon emission from gluon-gluon
scattering. Recently in Ref.~\cite{das10} a correction to the GB formula
for soft gluon emission was obtained. In present article we make an effort
to generalise the GB formula for soft gluon emission from gluon-gluon
scattering, {\it i.e.}, $gg\rightarrow ggg$ and find a more important
correction than it is found in Ref.~\cite{das10}. We also show that in an 
appropriate limit the generalisation reduces to the GB formula. 
We further note that the results for similar inelastic processes can be obtained in
a straightforward way by using our generalisation.

For the process, $gg\rightarrow ggg$, there are $25$ different topologies.
We note that $k_1$ and $k_2$ are  momenta of the gluons
in the entrance channel, $k_3$ and $k_4$ are those for exit channel gluons
whereas $k_5$ is that of the emitted gluon.
The invariant amplitude summed over all the final states and averaged over
initial states for such process can elegantly be 
written \cite{berends81} as:
\begin{eqnarray}
\left| {\cal M}_{gg \rightarrow ggg} \right | ^2
&=&\frac{1}{2}g^6\frac{N_c^3}{N_c^2-1}\frac{\cal{N}}{\cal{D}}
\lbrack(12345)+ (12354)\nonumber \\
&+&(12435)+(12453)+(12534)+(12543)\nonumber \\
&+&(13245)+(13254)+(13525)+(13424) \nonumber \\
&+&(14235) +(14325)  \rbrack \ ,  
\label{eq1}
\end{eqnarray}
where $N_c$ is the number of color, $g=\sqrt{4\pi\alpha_s}$ is the strong 
coupling, 
\begin{eqnarray}
{\cal N}&=&(k_1\cdot k_2)^4+(k_1\cdot k_3)^4+(k_1\cdot k_4)^4
+(k_1\cdot k_5)^4  \nonumber  \\ 
&&+(k_2\cdot k_3)^4+ (k_2\cdot k_4)^4+ (k_2\cdot k_5)^4+(k_3\cdot k_4)^4
 \nonumber  \\  
 &&+(k_3\cdot k_5)^4+(k_4\cdot k_5)^4,
\label{eq2}
\end{eqnarray}

\begin{eqnarray}
{\cal D}&=&(k_1\cdot k_2)(k_1\cdot k_3)(k_1\cdot k_4)(k_1\cdot k_5)
(k_2\cdot k_3)\nonumber \\ 
&&(k_2 \cdot k_4)
(k_2\cdot k_5)(k_3\cdot k_4)(k_3\cdot k_5)(k_4\cdot k_5) \ ,
\label{eq3}
\end{eqnarray}
and
\begin{equation}
(ijklm)=(k_i\cdot k_j)(k_j\cdot k_k)(k_k\cdot k_l)
(k_l\cdot k_m)(k_m\cdot k_i) \ .
\label{eq4}
\end{equation}

We now define the Mandelstram variables as
\begin{eqnarray}
 &&\hspace{-0.4cm}s=(k_1+k_2)^2\,, \hspace{0.4cm}  t= (k_1-k_3)^2\,,\hspace{0.3cm}   u=(k_1-k_4)^2,  \nonumber \\
 &&\hspace{-0.5cm}s^\prime=(k_3+k_4)^2\,, \hspace{0.2cm}
\ t^\prime=(k_2-k_4)^2\,,\hspace{0.2cm} u^\prime=(k_2-k_3)^2 , \nonumber 
\end{eqnarray}
and 
\begin{eqnarray}
&&\hspace{-0.4cm}k_1\cdot k_2=\frac{s}{2}\,, \hspace{0.9cm}k_3\cdot k_4=\frac{s^\prime}{2}\,, \hspace{0.75cm}  
k_1\cdot k_3=-\frac{t}{2}, \nonumber \\
&&\hspace{-0.4cm} k_2\cdot k_4=-\frac{t^\prime}{2}\,, \hspace{0.5cm} \ k_1\cdot k_4=-\frac{u}{2} 
\,, \hspace{0.5cm} k_2\cdot k_3=-\frac{u^\prime}{2}\,, \ \nonumber \\
&& k_1\cdot k_5=\frac{(s+t+u)}{2} \,, \hspace{0.5cm} 
k_2\cdot k_5=\frac{(s+t'+u')}{2}, \  \nonumber \\
&& k_3\cdot k_5=\frac{(s+t'+u)}{2}\,, \hspace{0.5cm} 
k_4\cdot k_5=\frac{(s+t+u')}{2} \, , 
\label{eq5}
\end{eqnarray}
with,
\begin{eqnarray}
 &&\hspace{-0.4cm} s+t+u+s'+t'+u'=0 \ . \nonumber 
\end{eqnarray}

Eq.~(\ref{eq1}) actually contains only twelve terms which are 
reduced from total 120 terms
due to symmetry~\cite{berends81}. Using (\ref{eq2}), (\ref{eq3})
and (\ref{eq4}) the first two
terms of (\ref{eq1}) can be expressed  as:
\begin{eqnarray}
\frac{1}{2}g^6\frac{N_c^3}{N_c^2-1}\frac{{\cal N}}
{(k_1\cdot k_3)(k_1\cdot k_4)(k_2\cdot k_4)(k_2\cdot k_5)
(k_3\cdot k_5)}  \nonumber \\
=\frac{1}{2}g^6\left(\frac{N_c^3}
{N_c^2-1}\right)\frac{8.4}{-t{t^\prime} u(s+{t^\prime}+{u^\prime})
(s+{t^\prime}+u)},
\hspace*{0.12in} \label{eq6}
\end{eqnarray}
and,
\begin{eqnarray}
\frac{1}{2}g^6\frac{N_c^3}{N_c^2-1}\frac{{\cal N}}
{(k_1\cdot k_3)(k_1\cdot k_5)(k_2\cdot k_4)(k_2\cdot k_5)
(k_3\cdot k_4)}   \nonumber \\
=\frac{1}{2}g^6\left(\frac{N_c^3}{N_c^2-1}\right)\frac{8.4}
{t{t^\prime} {s^\prime} (s+t+u)(s+{t^\prime}+{u^\prime})}
, \label{eq7}
\end{eqnarray}
After simplifying all the terms in this way, (\ref{eq1}) can be written as
\begin{eqnarray}
&&\left| {\cal M}_{gg \rightarrow ggg} \right|^2 =
16 g^6\frac{N_c^3}{N_c^2-1}{\cal N}  \, \nonumber \\
&&\times \left[ \frac{1}{s'(s+u+t)(s+u'+t')} \left ( \frac{1}{tt'}+\frac{1}{uu'}
\right ) \right.  \nonumber \\
&&\, \, - \frac{1}{t'(s+u+t)(s+u+t')} \left ( \frac{1}{ss'}+\frac{1}{uu'}
\right )   \nonumber \\
&&\, \, - \frac{1}{u'(s+u+t)(s+u'+t)} \left ( \frac{1}{tt'}+\frac{1}{ss'}
\right )   \nonumber \\
&&\, \, + \frac{1}{s(s+u+t')(s+u'+t)} \left ( \frac{1}{tt'}+\frac{1}{uu'}
\right )   \nonumber \\
&&\, \, - \frac{1}{u(s+u'+t')(s+u+t')} \left ( \frac{1}{tt'}+\frac{1}{ss'}
\right )   \nonumber \\
&&\, \,\left. - \frac{1}{t(s+u'+t')(s+u'+t)} \left ( \frac{1}{ss'}+\frac{1}{uu'}
\right )  \right ] 
\label{eq8}
\end{eqnarray}
and ${\cal N}$ can also be obtained as
\begin{eqnarray}
{\cal N}&=&\frac{1}{16}(s^4+t^4+u^4+{s^\prime}^4
+{t^\prime}^4+{u^\prime}^4)
  \nonumber \\
&&+\frac{1}{16}\left [ (s+t+u)^4
+(s+{t^\prime}+{u^\prime})^4+(s+{t^\prime}+u)^4 \right. 
\nonumber \\
&&\left.
+(s+t+{u^\prime})^4\right ]. 
\label{eq9}
\end{eqnarray}

Now for a soft gluon emission ($k_5\rightarrow 0$):
$t^\prime \rightarrow t, \ \ s^\prime \rightarrow s, \ \ 
u^\prime \rightarrow u$ and we can express the transverse component 
of the momentum of the emitted gluon in the centre of momentum frame of 
$k_1$ and $k_2$ as
\begin{eqnarray}
k_\perp^2 &=& \frac{4(k_1\cdot k_5)(k_2\cdot k_5)}{s}  
=\frac{(s+t+u)(s+u'+t')}{s}\nonumber \\
&=&\frac{(s+t+u)^2}{s} \ . 
\label{eq10}
\end{eqnarray}

Using (\ref{eq10}) in (\ref{eq8}) one can obtain a complete expression
for the three-body matrix element in terms of the two-body matrix element
and a infrared factor as
\begin{eqnarray}
\left| {\cal M}_{gg \rightarrow ggg} \right|^2=g^{2}
\left| {\cal M}_{gg \rightarrow gg} \right|^2
\frac{1}{k_\perp^{2}}\left[a_{1}+a_{2}\frac{t^4}{s^4}+a_{3
}\frac{k_\perp^{4}}{s^2}\right] \nonumber \\
\times\left[\left(a_{4}+a_{5}\frac{t}{s}+a_{6}\frac{t^{2}}{s^{2}}
\right)/\left(a_{7}+a_{8}\frac{t^{2}}{s^{2}}+a_{9}
\frac{t^{3}}{s^{3}}\right)\right], \hspace*{0.2in}
\label{eq11}
\end{eqnarray}
where the full two-body matrix element~\cite{xiong94,kcr77} is given as
\begin{equation}
\left| {\cal M}_{gg \rightarrow gg} \right|^2
=\frac{9}{2}g^4 \frac{s^2}{t^2}\left[ a_7
+a_8\frac{t^2}{s^2}+a_9
\frac{t^3}{s^3}\right],  
\label{eq12}
\end{equation}
and various coefficients are
\begin{eqnarray}
&&a_{1}=3+3\frac{u^4}{s^4}, \ \ \  
a_{2}=3, \ \ \   
a_{3}=6, \ \ \
a_{4}=1-\frac{s}{u},  
\nonumber \\
&&a_{5}=-1-\frac{s^2}{u^2}, \ \ 
a_{6}=\frac{s^2}{u^2}-\frac{s}{u}, \ \  
a_{7}=-\frac{u}{s}, \ \ 
a_{8}=3 , 
\nonumber \\
&&a_{9}=-\left (\frac{u}{s}+\frac{s^2}{u^2}\right ) . 
\hspace*{0.3in} \label{eq13}
\end{eqnarray}

Furthermore, eliminating $u$ using (\ref{eq10}) and keeping 
terms upto ${\cal O}(1/k_\bot^2)$ and  ${\cal O}(t^3/s^3)$ one can
write (\ref{eq11}) as 
\begin{eqnarray}
\left| {\cal M}_{gg \rightarrow ggg} \right|^2&=&
12g^{2}\left| {\cal M}_{gg \rightarrow gg} \right|^2_{GB}
\frac{1}{k_\perp^{2}}\left[1+\frac{t}{2s}\right.\nonumber \\
&&\left.+\frac{5}{2}\frac{t^2}{s^2}-\frac{t^{3}}{s^{3}}
+{\cal O}\left(\frac{t^{4}}{s^{4}}\right)\right],  
\label{eq14} 
\end{eqnarray}
where
\begin{eqnarray}
\left| {\cal M}_{gg \rightarrow gg} \right|^2_{GB}&=& \frac{9}{2}g^4
\frac{s^2}{t^2} \ .
\label{eq14a} 
\end{eqnarray}
The above equation (\ref{eq14}) is a convergent series of $t/s$ as 
can be seen below. It is now clearly decomposed in two factors: one is 
associated with the
$2\rightarrow 2$ process used by Gunion and Bertsch~\cite{gunion82}
whereas the other one is the generalisation
of the infrared factor for the emission of soft quanta. As we will see below, 
the first term in (\ref{eq14}) will lead to the GB formula~\cite{gunion82} for 
the gluon multiplicity distribution in an appropriate 
limit $\frac{1}{k_\bot^2}\approx \frac{1}{k_\bot^2} 
\frac{q_\bot^2}{(k_\bot-q_\bot)^2}$, where $t=q_\bot^2>> k_\bot^2$
($q_\bot$ is the transverse component of the momentum transfer). If
the emitted gluon is much softer than others it can then be 
regulated~\cite{biro93,wong96} by the Debye screening mass, $m_D$. 
On the other hand the second, the third and the fourth terms, 
respectively, would correspond to the correction terms over the GB 
term in the same spirit.
We also note that the first term can also be obtained by just using 
the scalar QCD approximation in $2\rightarrow 3$ process.

If one considers only $a_{1},a_{4},a_{6},a_{7}$ in (\ref{eq11}) 
and substitute $u=-s$, one will end up with the expression derived in 
Ref.~\cite{das10} as
\begin{eqnarray}
\left| {\cal M}_{gg \rightarrow ggg} \right|^2=
12g^{2}\left| {\cal M}_{gg \rightarrow gg} \right|^2_{GB}
\frac{1}{k_\perp^{2}}
\left(1+\frac{t^{2}}{s^{2}}\right), 
\label{eq15} 
\end{eqnarray}
which misses the leading order correction term of ${\cal O}(t/s)$ 
and also the right coefficient  in the next order, ${\cal O}(t^2/s^2)$.

Using (\ref{eq14}) and following Refs.~\cite{gunion82,wong96}, it is 
now very straight forward to obtain the soft gluon
multiplicity distribution in the midrapidity region as 
\begin{eqnarray}
\frac{dn_g}{d\eta dk_\perp^2}\,&=&\,
\left[\frac{dn_g}{d\eta dk_\perp^2}\right]_{GB}\,
\left(1+\frac{(q_\perp^2+m_D^2)}{2s}\right.\nonumber \\
&&\left.+\frac{5}{2}\frac{(q_\perp^2+m_D^2)^2}{s^2}
-\frac{(q_\perp^2+m_D^2)^3}{s^3}\right),
\label{eq16}
\end{eqnarray}
where the GB formula can be obtained using 
(\ref{eq14a}) as 
\begin{eqnarray}
\left[\frac{dn_g}{d\eta dk_\perp^2}\right]_{GB}&=&
\frac{C_A\alpha_s}{\pi^2}
\frac{q_\perp^2}{k_\perp^2\lbrack(k_\perp-q_\perp)^2+m_D^2\rbrack},
\label{eq17}
\end{eqnarray}
with the Casimir factor $C_A=3$ and the Debye screening mass 
$m_D=gT$. We note that the first term in (\ref{eq16}) would correspond to
GB formula as obtained in Ref.~\cite{gunion82}. It is also worth mentioning that
this first term can be obtained by just using the scalar QCD approximation.

Now the average value of the momentum transfer squared can be obtained as
\begin{eqnarray}
\langle q_\bot^2\rangle =\left (  {\int\limits_{m_D^2}^{s/4} dt \ t \ 
\frac{d\sigma}{dt}}\right ) / \left (
{\int\limits_{m_D^2}^{s/4} dt \  
\frac{d\sigma}{dt}} \right ) ,
\label{eq18}
\end{eqnarray}
where the maximum value of $\alpha_s$ is restricted by $s\ge4m_D^2$
along with $s=\langle s\rangle =18T^2$.

\begin{figure}[t]
\includegraphics[height=2.8in,width=3.3in]{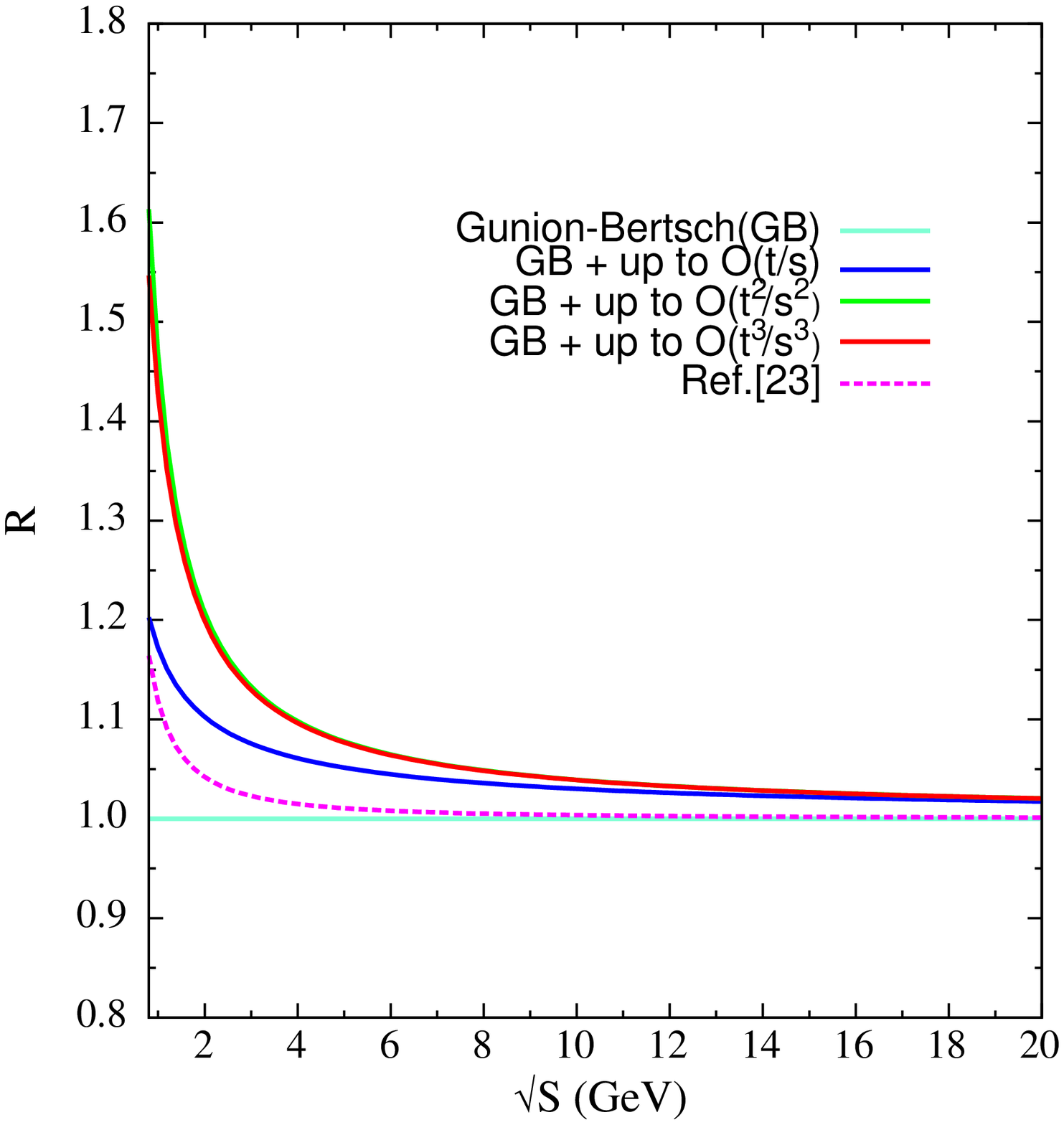}
\caption{The ratio 
$R=\frac{dn_g}{d\eta dk_\bot^2}/\left.\frac{dn_g}{d\eta dk_\bot^2}\right|_{GB}$ 
 as a function of $\sqrt s$ at the center of momentum 
frame of gluon-gluon scattering at $T=200$ MeV and $\alpha_s=0.3$. }
\label{ratio_s}
\end{figure}

\begin{figure}[t]
\includegraphics[height=2.8in,width=3.3in]{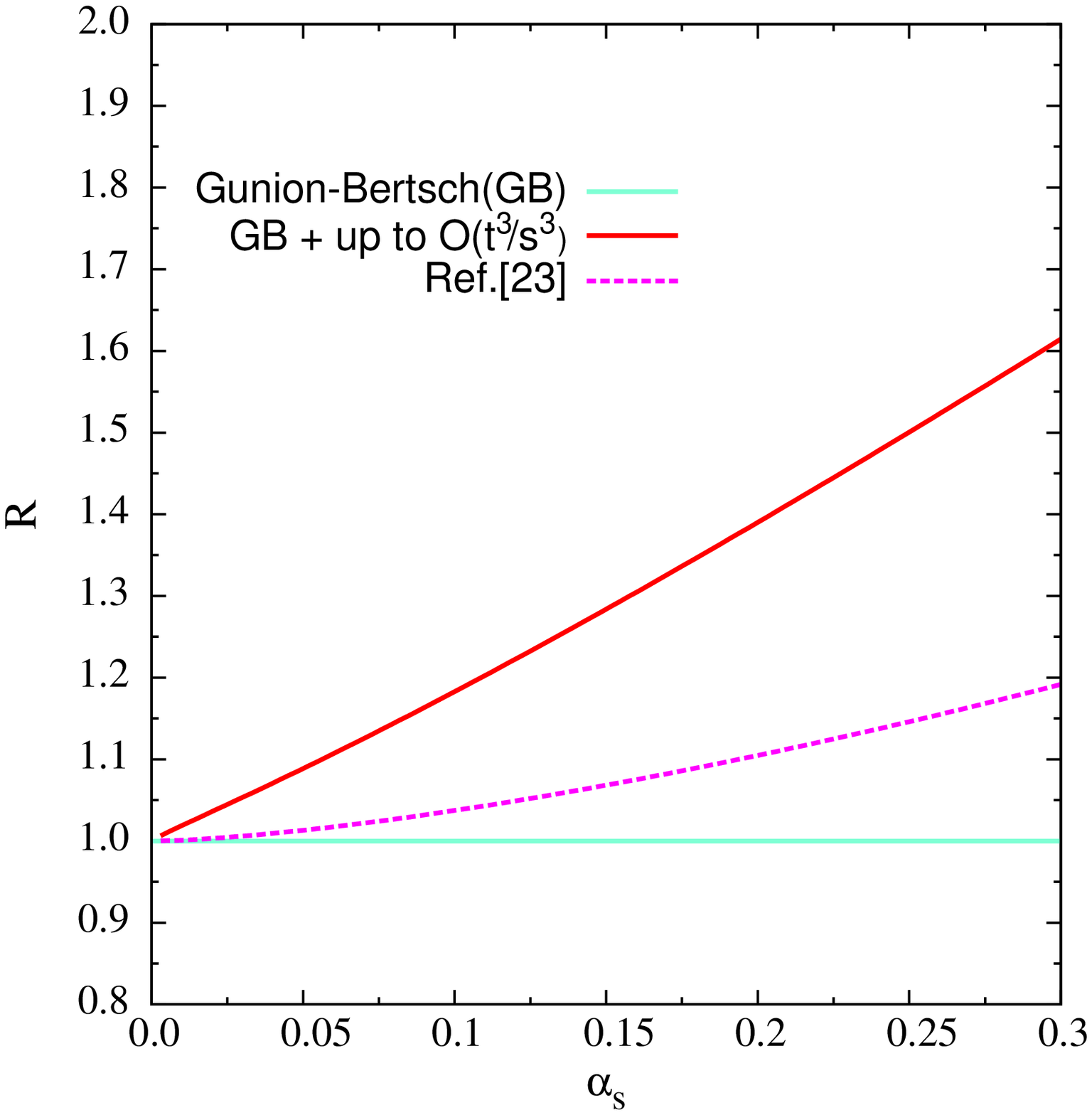}
\caption{The ratio 
$R=\frac{dn_g}{d\eta dk_\bot^2}/\left.\frac{dn_g}{d\eta dk_\bot^2}\right|_{GB}$ 
as a function of strong coupling, $\alpha_s$. }
\label{ratio_g}
\end{figure}

In Fig.~\ref{ratio_s} the ratio 
$R=\frac{dn_g}{d\eta dk_\bot^2}/\left.\frac{dn_g}{d\eta dk_\bot^2}\right|_{GB}$ 
is plotted with the center of momentum
energy, $\sqrt s$  of the gluon-gluon scattering for $T=200$ MeV 
and $\alpha_s=0.3$. With these values the lower limit of $\sqrt s\sim 0.8$GeV 
is restricted by the relation $s\ge 4m_D^2$. As can be seen from 
Fig.~\ref{ratio_s} that (\ref{eq14}) is a convergent series in $t/s$ and it
also indicates a significant improvement compared to that of Ref.~\cite{das10}.
With the increase of $\sqrt s$ 
the correction terms decreases and the ratio approaches the GB formula for 
very large value of $\sqrt s$. On the other hand the first and second order
corrections are significant when $\sqrt s \le 6$ GeV. 

In Fig.~\ref{ratio_g} the ratio
$R=\frac{dn_g}{d\eta dk_\bot^2}/\left.\frac{dn_g}{d\eta dk_\bot^2}\right|_{GB}$ 
is plotted as a function of the strong coupling $\alpha_s$. 
We note that the maximum value of $\alpha_s$ is restricted by $s\ge 4m_D^2$.
As can be seen the correction terms become important with the increase of 
$\alpha_s$.  At $\alpha_s=0.3$, the correction up to ${\cal O}(t^3/s^3)$ 
is around $50\%$ over the usual GB formula. We have also compared our results 
with that of Ref.~\cite{das10}, which shows a significant improvement
in the range of $\alpha_s$ displayed in Fig.~\ref{ratio_g}.

\begin{figure}[t]
\vspace*{0.3in}
\includegraphics[height=2.8in,width=3.3in]{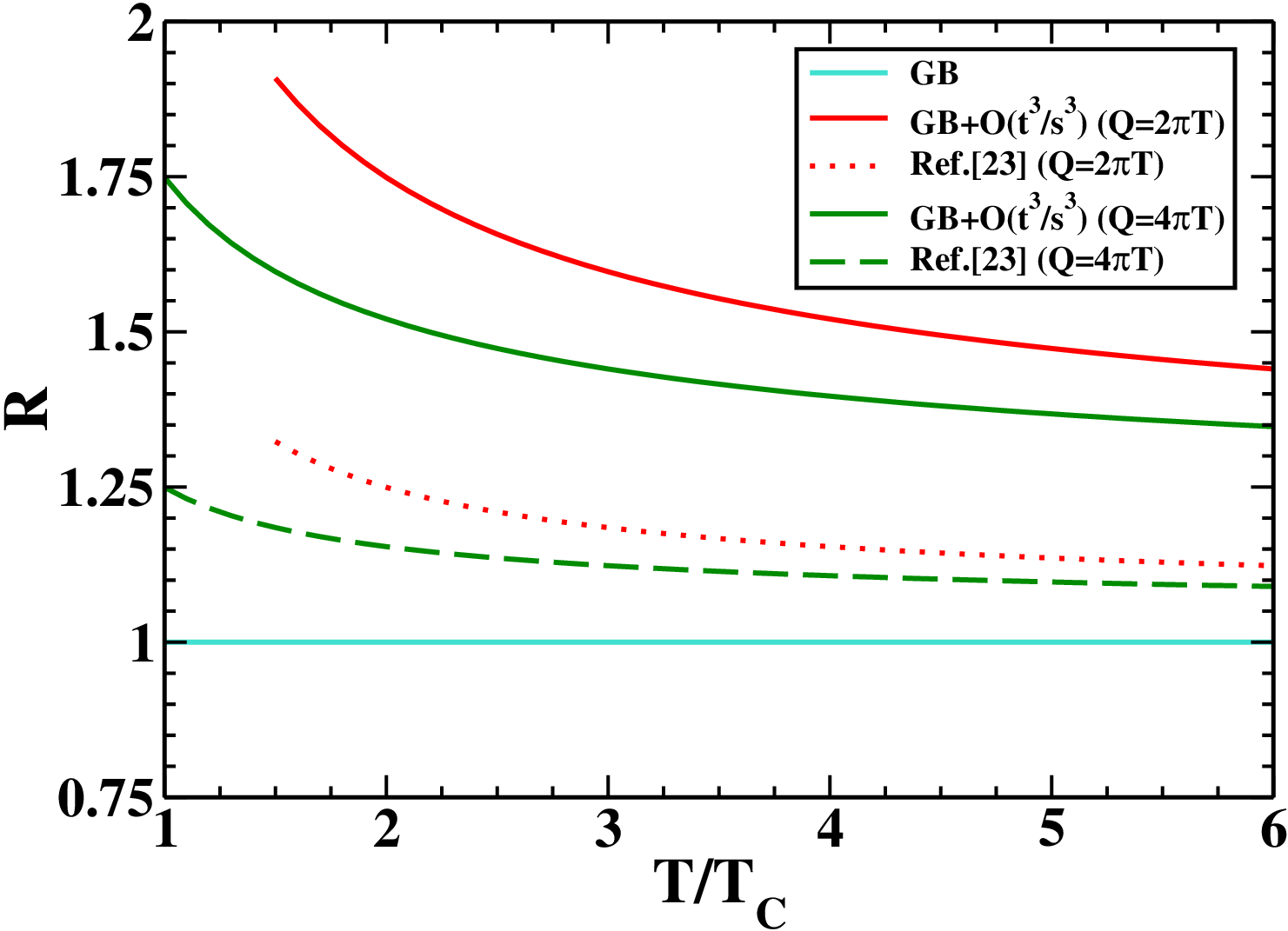}
\caption{The ratio 
$R=\frac{dn_g}{d\eta dk_\bot^2}/\left.\frac{dn_g}{d\eta dk_\bot^2}\right|_{GB}$ 
 up to ${\cal O}(t^3/s^3)$ is displayed 
as a function of $T/T_c$ for temperature dependent $\alpha_s$ with
two momentum scales $2\pi T$ (red curve) and $4\pi T$ (green curve). 
The corresponding results from Ref.~\cite{das10} are represented by
dotted (red) and  dashed (green) curves.} 
\label{ratio_T}
\end{figure}

Fig.~\ref{ratio_T} displays the ratio $R$ as a function of $T$ in the units
of the critical temperature $T_c$. Here we have used the temperature
dependent $\alpha_s$ from Ref.~\cite{chakraborty02} with two momentum scale
$Q=2\pi T$ (red curve) and $4\pi T$ (green curve). We note that 
for the momentum scale $Q=2\pi T$ the value of the coupling, 
$\alpha_s > 0.4$ for $T/T_C < 1.5$, which is really a very high for
any practical purposes.  On the other hand for 
$Q=4\pi T$, the values of coupling lie in the domain $0.2\le\alpha_s\le 0.35$ 
for the scaled temperature range, $1\le T/T_C\le 6$. Nonetheless, the value
of $\alpha_s$ and thus the lower bound of $T/T_C$  would again be 
restricted by the relation $s\ge 4m_D^2$.  As can be seen in 
Fig.~\ref{ratio_T}, there is a sizeable contribution ($\ge 40\%$) up to 
${\cal O}(t^3/s^3)$ over the GB formula and a correction ($\geq 30\%$) 
compared to Ref.~\cite{das10} in the temperature range $1\le T/T_C\le 5$. 
This correction over the GB formula would be very important 
at the temperature domain relevant for the heavy-ion collisions.

In summary, we have obtained a generalisation of the GB formula for soft gluon
emission in a process $gg\rightarrow ggg$. We found that the correction terms
are important at various physical domains of temperature, 
coupling constant and the energy of gluon-gluon scattering. This generalisation
will be very apt for the phenomenology ({\it viz.}, hot glue scenario,
chemical equilibration of gluons, partonic matter viscosity, 
radiative energy-loss of energetic 
partons and jet quenching) of heavy-ion collisions and would improve the 
present understanding on various phenomena in this area.

\vspace*{0.3in}

\begin{acknowledgments}
We thank to Jorge Noronha for useful discussions. RA is thankful to Swapan Majhi
for various discussions and help during the course of this work. 
This work was partly supported by the Helmholtz International
Centre for FAIR within the framework of LOEWE (Landes-Offensive zur Entwicklung
Wissenschaftlich-{\"o}konomischer Exzellenz) program launched by the state of
Hesse, Germany.
\end{acknowledgments}

\end{document}